\documentclass[onecolumn,12pt]{article}
\usepackage{times}
\usepackage{graphicx}
\usepackage{natbib}
\usepackage[margin=1.0in]{geometry}

\usepackage[utf8]{inputenc}
\usepackage{enumitem,amssymb}
\usepackage{ragged2e}
\newlist{thematic}{itemize}{8}
\setlist[thematic]{label=$\square$}
\usepackage{pifont}

\usepackage{graphicx} 
\usepackage{longtable} 
\usepackage{rotating}
\usepackage{tabularx}
\usepackage{hyperref}
\usepackage{easylist}
\usepackage{amsmath}
\usepackage{bm}
\usepackage{array} 
\usepackage{amssymb,amsmath} 
\usepackage{sidecap}

\newcommand{\kep}{\mbox{\textit{Kepler}}}
\newcommand{\logg}{\mbox{$\log g$}}
\newcommand{\tess}{{\it TESS}}

\usepackage{titlesec}
\titlespacing{\section}{0pt}{3ex}{1ex}

\begin{document}
\raggedright
\huge
Astro2020 Science White Paper \linebreak

Stellar Physics and Galactic Archeology using Asteroseismology in the 2020's
 \linebreak
\normalsize

\noindent \textbf{Thematic Areas:} \hspace*{60pt} $\boxtimes$ Planetary Systems \hspace*{10pt} $\square$ Star and Planet Formation \hspace*{20pt}\linebreak
$\square$ Formation and Evolution of Compact Objects \hspace*{31pt} $\square$ Cosmology and Fundamental Physics \linebreak
  $\boxtimes$  Stars and Stellar Evolution \hspace*{1pt} $\boxtimes$ Resolved Stellar Populations and their Environments \hspace*{40pt} \linebreak
  $\square$    Galaxy Evolution   \hspace*{45pt} $\square$             Multi-Messenger Astronomy and Astrophysics \hspace*{65pt} \linebreak
  
\textbf{Principal Author:}

Name: Daniel Huber
 \linebreak						
Institution: Institute for Astronomy, University of Hawai`i
 \linebreak
Email: huberd@hawaii.edu
 \linebreak
Phone: 808-956-8573 
 \linebreak
 
\textbf{Co-authors:} 
\linebreak
Sarbani Basu, Yale University \linebreak
Paul Beck, Instituto de Astrofisica de Canarias \linebreak
Timothy R.\ Bedding, University of Sydney \linebreak
Derek Buzasi, Florida Gulf Coast University \linebreak
Matteo Cantiello, Center for Computational Astrophysics \linebreak
William J.\ Chaplin, University of Birmingham \linebreak
Jessie L.\ Christiansen, Caltech/IPAC-NExScI \linebreak
Katia Cunha, University of Arizona \linebreak
Ricky Egeland, NCAR High Altitude Observatory \linebreak
Jim Fuller, California Institute of Technology  \linebreak
Rafael A. Garc\'\i a, Astronomy Division, CEA \linebreak
Douglas R.\ Gies, Georgia State University \linebreak
Joyce Guzik, Los Alamos National Laboratory \linebreak 
Saskia Hekker, Max Planck Institute for Solar System Research \linebreak
JJ Hermes, Boston University \linebreak
Jason Jackiewicz, New Mexico State University \linebreak
Jennifer Johnson, Ohio State University \linebreak
Steve Kawaler, Iowa State University \linebreak
Travis Metcalfe, Space Science Institute \linebreak
Benoit Mosser, Observatoire de Paris  \linebreak
Melissa Ness, Columbia University / Flatiron Institute  \linebreak
Marc Pinsonneault, Ohio State University \linebreak
Anthony L. Piro, Carnegie Observatories\linebreak
Victor Silva Aguirre, Aarhus University \linebreak
David Soderblom, Space Telescope Science Institute \linebreak
Keivan Stassun, Vanderbilt University \linebreak
Jamie Tayar, University of Hawai`i, Hubble Fellow \linebreak
Theo ten Brummelaar, Georgia State University \linebreak
Rachael Roettenbacher, Yale University \linebreak
Jennifer Sobeck, University of Washington \linebreak
Regner Trampedach, Space Science Institute \linebreak
Gerard van Belle, Lowell Observatory \linebreak
Jennifer van Saders, University of Hawai`i \linebreak
Dennis Stello, University of New South Wales \linebreak

\textbf{Abstract:}

\justifying

\noindent
Asteroseismology is the only observational tool in astronomy that can probe the interiors of stars, and is a benchmark method for deriving fundamental properties of stars and exoplanets. Over the coming decade, space-based and ground-based observations will provide a several order of magnitude increase of solar-like oscillators, as well as a dramatic increase in the number and quality of classical pulsator observations, providing unprecedented possibilities to study stellar physics and galactic stellar populations. In this white paper, we describe key science questions and necessary facilities to continue the asteroseismology revolution into the 2020's.

\pagebreak
\section{Asteroseismology in the 2020's}

\noindent
Asteroseismology -- the study of stellar oscillations -- is one of the most powerful tools to probe the interiors of stars across the H-R diagram (Figure \ref{HRfigure}). The field has undergone a revolution over the past decade driven by space-based photometry provided by CoRoT \citep{baglin06b} and \kep\ \citep{borucki10}. In particular, the detection of oscillations in thousands of low-mass stars has led to major breakthroughs in stellar astrophysics, such as the discovery of rapidly rotating cores in subgiants and red giants, as well as the systematic measurement of stellar masses, radii and ages \citep[see][]{chaplin13a}. Asteroseismology has also become the ``gold standard'' for calibrating more indirect methods to determine stellar parameters such as surface gravity (\logg) from spectroscopy \citep{petigura17b} and stellar granulation \citep{bastien13,kallinger16,bugnet2018,pande2018}, and age from rotation periods \citep[gyrochronology, e.g.][]{vansaders16} or magnetic activity \citep{metcalfe19}. 


The asteroseismology revolution is set to continue over the coming decade. The recently launched NASA \tess\ Mission \citep{ricker14} is expected to detect oscillations in thousands of main-sequence and subgiant stars \citep{schofield19}, an order of magnitude increase over \kep. Light curves from \tess\ full-frame images are expected to yield hundreds of thousands of oscillating red giants over the coming years. Further into the next decade,  H-band light curves by the WFIRST microlensing survey are expected to yield over one million oscillating red giants in the galactic bulge \citep{gould15}, while the the European-led PLATO mission will yield detections in nearly 100,000 dwarfs and subgiants, as well as potentially millions of red giants \citep{rauer14,miglio17,mosser19}. 

Combined, the sample of solar-like oscillators is expected to increase by several orders of magnitude over the coming decade (Figure \ref{HRfigure}), providing unprecedented datasets for stellar astrophysics, exoplanet science, and galactic astronomy. In this white paper we discuss key science questions and challenges for asteroseismology in the 2020's.

\section{Stellar Physics}

\noindent
Stellar evolution theory forms one of the backbones of modern astronomy. Guided by measurements of fundamental stellar properties from observations, stellar models affect a wide range of fields in astrophysics by underpinning stellar population synthesis models that are used to study galaxy evolution and cosmology. On a smaller scale, properties of exoplanets critically depend on the characteristics of host stars, which frequently dominate the error budget \citep{huber18}.

Despite significant advances, many problems in stellar physics remain unsolved. Uncertainties in the description of convective core overshooting leads to different main-sequence lifetimes for stars more massive than the Sun, while the general treatment of convective energy transport leads to different predictions of radii for lower mass stars. For red giants, major uncertainties include interior angular momentum transport and mass loss, leading to systematic errors in ages of up 50\% \citep{casagrande14} and significantly affecting the late stages of stellar evolution.

\begin{figure}
\begin{center}
\resizebox{\hsize}{!}{\includegraphics{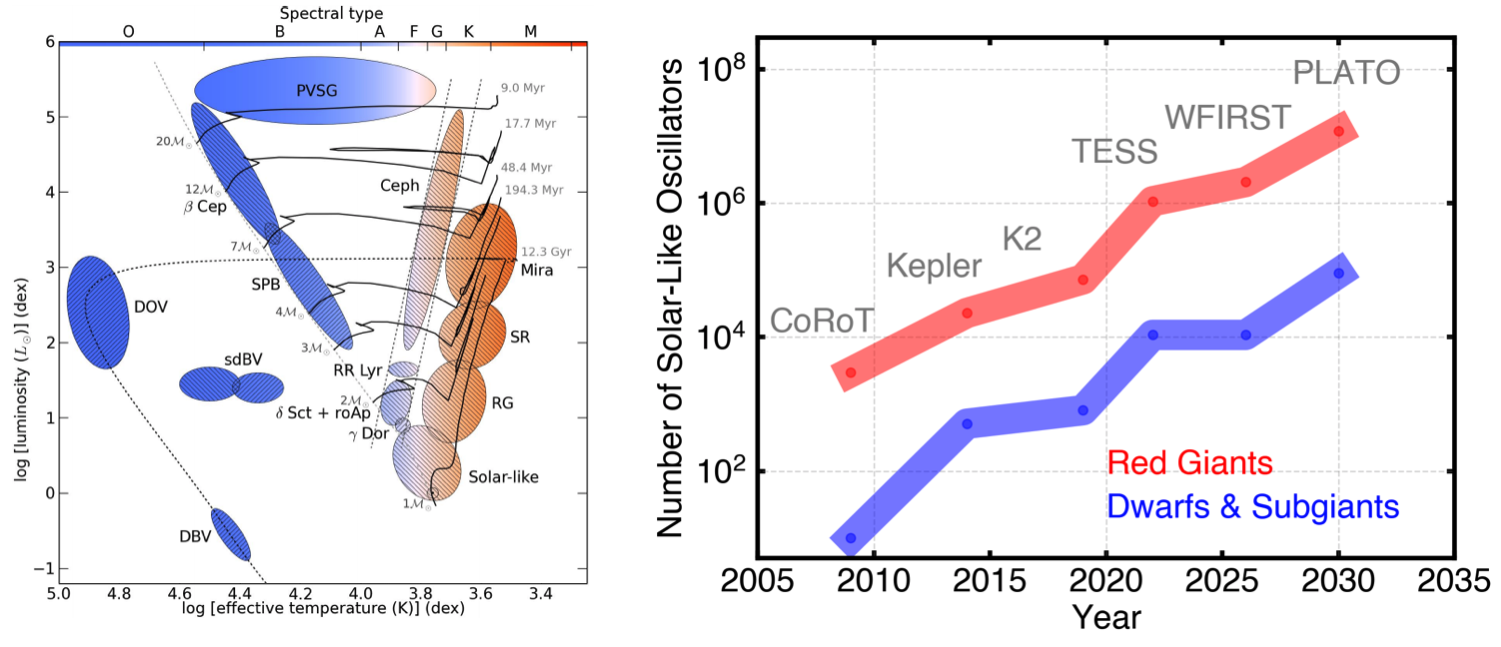}}
\caption{Left: Pulsating stars across the HR diagram (from P. Papics, based on a figure by  J.~Christensen-Dalsgaard). Right: Yield of solar-like oscillators discovered by space-based telescopes, separated into dwarfs and subgiants (which require faster than 30 minute sampling) and red giants. The several orders of magnitude yield increase in the 2020's will provide unprecedented opportunities for stellar physics, exoplanet science and galactic astronomy.}
\label{HRfigure}
\end{center}
\end{figure}



Asteroseismology is a unique tool to address open problems in stellar modeling. Oscillation frequencies are sensitive to the interior sound speed profile, placing constraints on stellar structure such as the depth of the convective envelopes \citep{mazumdar14}, the presence of convective cores \citep{silva13}, magnetic fields \citep{fuller15,stello16}, physical processes such as interior angular momentum transport \citep{beck12, deheuvels14}, and fundamental stellar properties such as masses and ages \citep{metcalfe12, creevey17}. While our discussion here primarily concerns stochastically-excited solar-like stars, many of these processes can also be probed in classical pulsators throughout the H-R diagram such as Miras, Cepheids, RR Lyrae and Delta Scuti stars \citep{molnar18,kolenberg18}, providing an unparalleled window into not only internal stellar physics, but also galactic astronomy and -- through their contributions to understanding the distance ladder --  to cosmology (Figure \ref{HRfigure}).


A critical ingredient for asteroseismology are independent constraints on fundamental stellar properties to break parameter degeneracies \citep{cunha07}. For example, helium abundance, stellar mass and radius are strongly correlated based on the measurements of oscillation frequencies alone \citep{silva17}. Furthermore, effective temperatures are required to place stars onto a model grid. 
The solution to overcome this challenge is long-baseline interferometry, which allows measurements of stellar angular diameters. Combined with parallaxes and bolometric fluxes, interferometry allows the only nearly-model-independent measurements of stellar radii \citep{boyajian13}, including effects of rotation \citep{maestro13,jones16}. Such fundamental radius constraints are not possible with \textit{Gaia} parallaxes alone.

\begin{figure}
\begin{center}
\resizebox{\hsize}{!}{\includegraphics{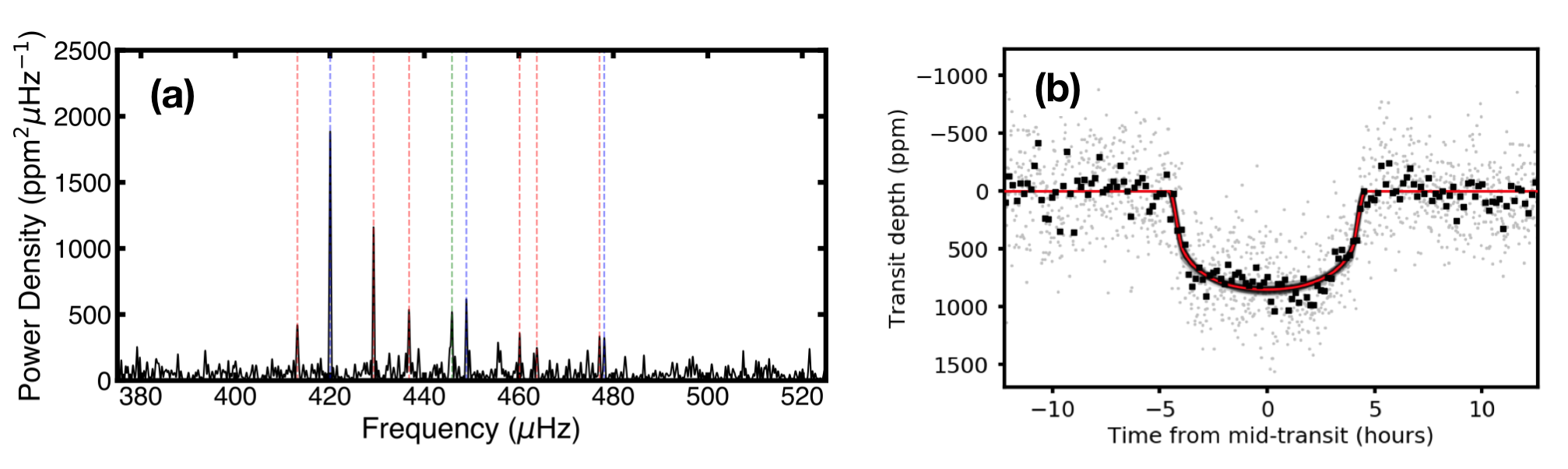}}
\caption{Power spectrum (panel a) and phase-folded transit light curve (panel b) of the bright subgiant TOI-197 ($V=8.1$), the first exoplanet host with solar-like oscillations detected by \tess. 
Stars like TOI-197 will be within reach of next-generation interferometric instruments, providing unprecedented constraints on the star and its orbiting planet. From \citet{huber19}.}
\label{toi197}
\end{center}
\end{figure}

So far, the overlap between asteroseismology and interferometry has been limited to a handful of the brightest stars observed by \kep\ \citep{huber12, white13}, which will dramatically change with the brighter stars observed by \tess\ and PLATO. 
\textbf{Maintaining and expanding the leading US capabilities in optical long-baseline interferometry, such as the Center for High Angular Resolution (CHARA) Array \citep{brummelaar04}, the Navy Precision Optical Interferometer \citep[NPOI,][]{Armstrong2013JAI.....240002A} and Magdalena Ridge Observatory Interferometer \citep[MROI,][]{creech10}, will thus be critical to capitalize on the potential of asteroseismology to probe stellar physics in the upcoming decade.} Indeed, \tess\ has already started to deliver detections in bright oscillating exoplanet host stars, which will be within reach of next generation interferometric instruments (Figure \ref{toi197}).

The combination of bolometric fluxes with angular diameters also allows the fundamental measurement of effective temperatures. Bolometric fluxes inferred by combining broadband photometry with model atmospheres (SED fitting) are often plagued by systematic errors from photometric zeropoints \citep{mann15}, and space-based measurements are rare. As a result, effective temperature scales are still plagued by systematic errors \citep[e.g.][]{casagrande14c}, limiting our capacity for precision stellar astrophysics in the era of space-based photometry \citep{huber16b}. \textbf{Spectrophotometry from SPHEREx, combined with \textit{Gaia} and interferometry, will establish a definite effective temperature scale for stars across the H-R diagram, a critical bottleneck in stellar physics.} Combined, fundamental measurements of stellar temperature, radii and oscillation frequencies will provide unprecedented advances in our understanding of stellar physics over the coming decade, and refine the fundamental physics underlying our models of stars that are widely applied to frontier efforts in both 
extragalactic and exoplanet science.

\section{Galactic Archeology using Oscillating Red Giants}


\noindent
The study of the chemo-dynamical history of stellar populations in our galaxy (galactic archeology) is one of the most rapidly evolving fields in stellar astrophysics. Driven by large spectroscopic surveys such as  RAVE, Gaia-ESO, SDSS/APOGEE, LAMOST, and GALAH and kinematics from \textit{Gaia}, galactic archeology is starting to yield spectacular insights into stellar populations in our galaxy \citep[e.g.][]{hayden15,silva18,hawthorn19}.

A particularly powerful synergy is the combination of chemical abundances from spectroscopy with stellar masses from asteroseismology, allowing the determination of ages for red giant stars which serve as powerful probes of galactic stellar populations \citep{miglio12}. Indeed, the systematic combination of spectroscopic surveys with asteroseismology for the \kep\ field through the APOKASC collabration \citep{pinsonneault18} has led to intriguing discoveries such as a population of young stars with enhanced $\alpha$ abundances \citep{martig15}, and derived calibrations have enabled the systematic determination of ages for hundreds of thousands of stars throughout the galaxy \citep{ness16,ho17}.

\begin{SCfigure}
\resizebox{10cm}{!}{\includegraphics{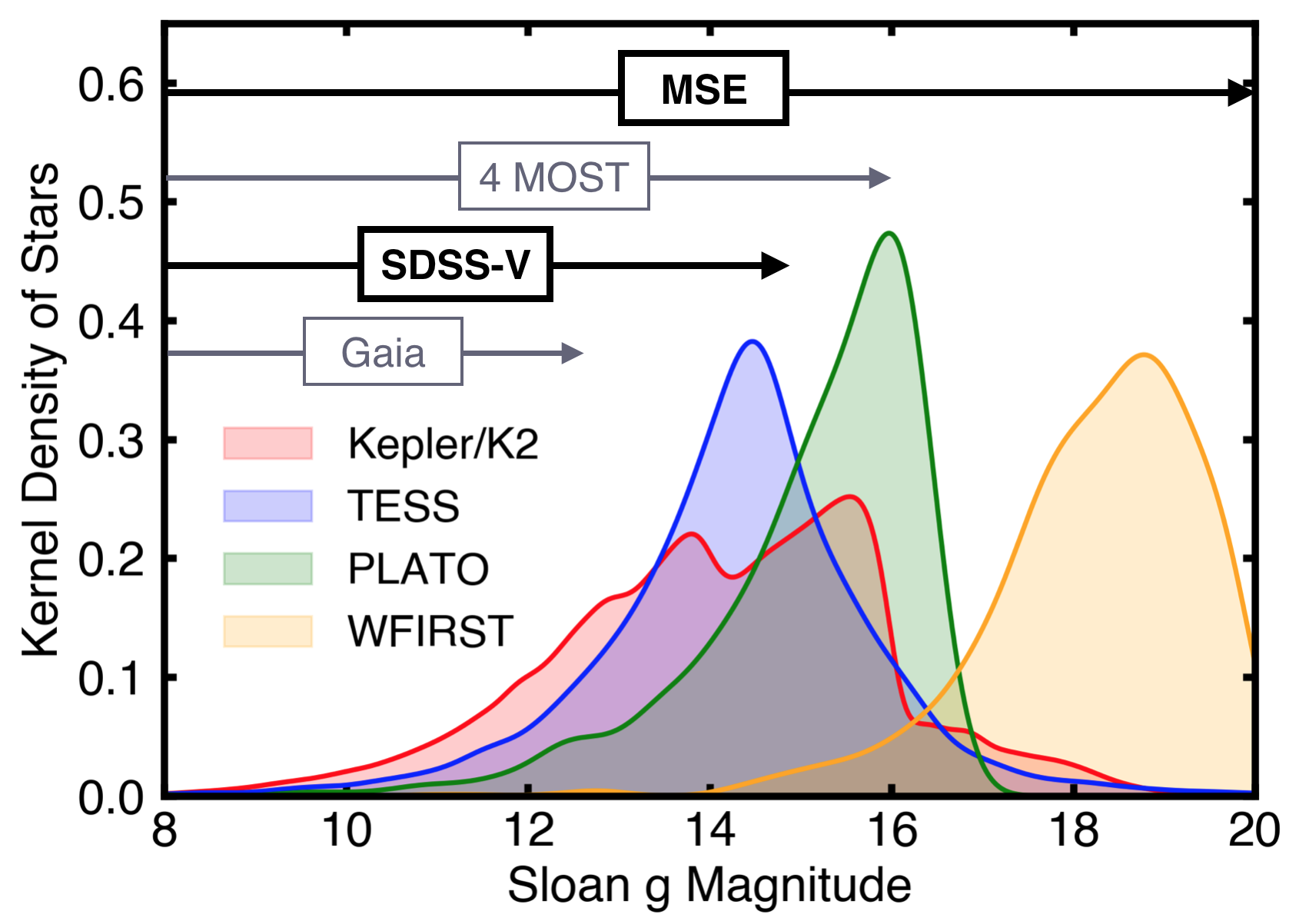}}
\caption{Sloan g-magnitudes for stars observed by Kepler/K2 \citep[red,][]{huber16} and predicted yields with photometric precision $<1$\,mmag\,hr$^{-1}$ from \tess\ \citep[blue,][]{stassun18}, a typical PLATO field \citep[green,][]{rauer14}, and the WFIRST microlensing campaign \citep[yellow,][]{gould15}. Sensitivity limits of MOS facilities with $R>20000$ spectroscopy covering $>2\,\pi$ of the sky are indicated. Black lines highlight planned US-led surveys in the 2020's.}
\label{ga}
\end{SCfigure}

The enormous increase in asteroseismic detections of red giants in the coming decade will  increase the reach of asteroseismology, and thus the need for spectroscopic follow-up observations. In the Kepler/K2 era, asteroseismic samples were small enough to obtain spectroscopic follow through single-object spectrographs for dwarfs \& subgiants \citep[e.g.][]{buchhave15} or existing multi-object spectrograph surveys such as SDSS-IV/APOGEE. \textbf{Massively multiplexed, large scale surveys  such as SDSS-V and the Maunakea Spectroscopic Explorer (MSE) are critical to obtain spectra of millions of stars for which we expect to obtain high-precision space-based photometry over the coming decades (Figure \ref{ga}).} Such surveys will not only benefit asteroseismology, but also a large range of other stellar astrophysics investigations such as stellar rotation, flares, the characterization of exoplanet host stars, and the detection and characterization of a wide range of binary systems through multi-epoch spectroscopy \citep{bergemann19}.

\section{Asteroseismology of Cool Dwarfs using Radial Velocities}

\noindent
Main sequence stars cooler than the Sun are ubiquitous in our galaxy, yet our understanding of their structure and evolution remains one of the most challenging problems in stellar evolution theory. In particular, stellar models systematically underestimate radii of K--M dwarfs at fixed temperature by up to 20\%\ compared to empirical radii from interferometry and eclipsing binaries \citep{kraus11,boyajian12}. While theories such as convective suppression by close binaries or magnetic fields can reduce the discrepancy \citep{feiden13}, a universal explanation remains elusive. This is particularly troublesome because these stars have become a primary focus for exoplanet science: both the NASA \tess\ Mission and JWST aim to characterize potentially habitable planets around cool dwarfs, as well as hotter planets around FGK stars.

\begin{SCfigure}
\resizebox{11cm}{!}{\includegraphics{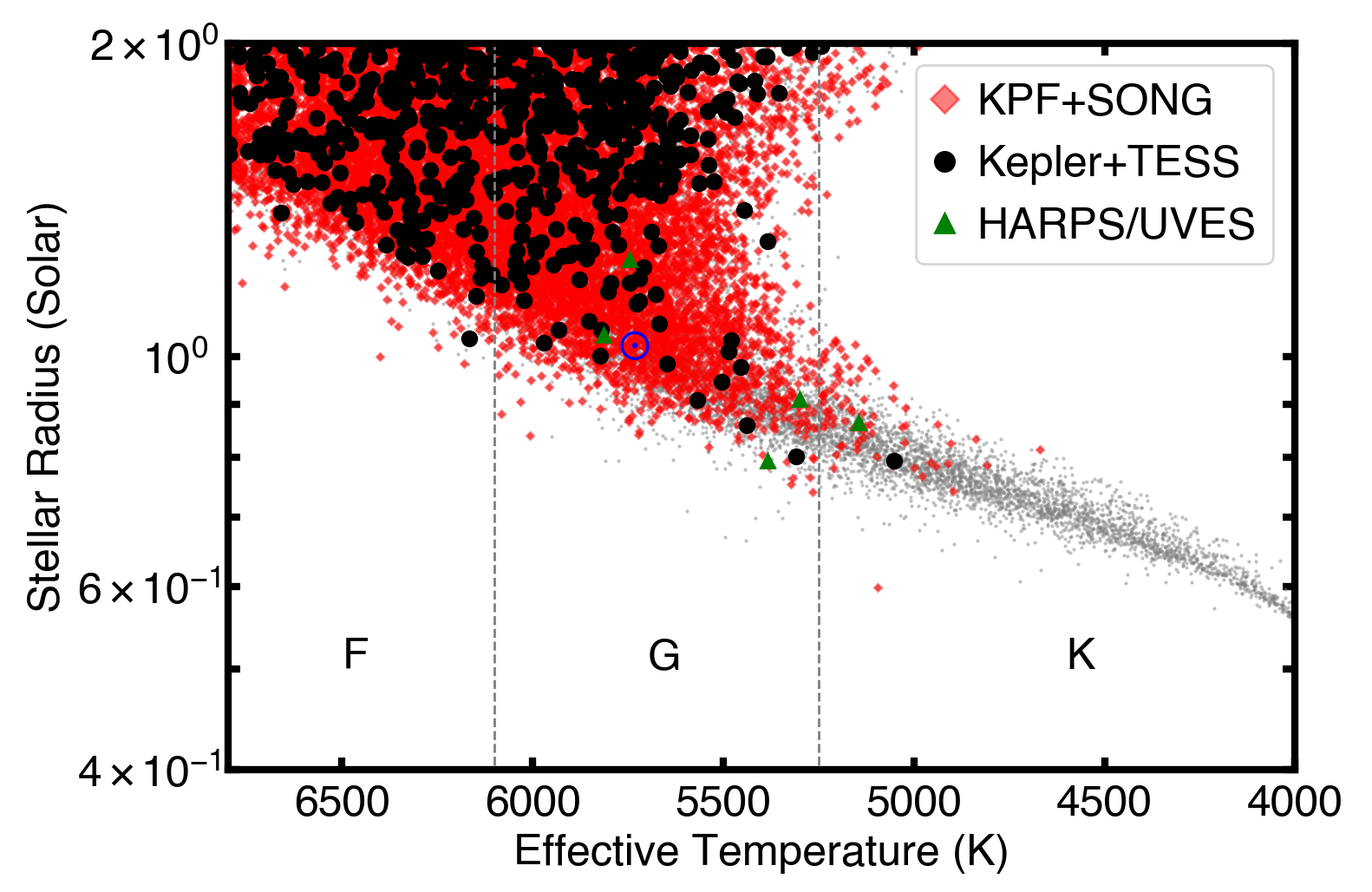}}
\caption{\textit{Gaia}-derived H-R diagram (grey) highlighting asteroseismic detections from space (Kepler \& \tess, black) and stars within the sensitivity limits of ground-based radial-velocity (RV) observations (KPF \& SONG, red). Next generation RV facilities such as SONG and KPF will allow the first systematic application of asteroseismology to stars cooler than the Sun.}
\label{cooldwarfs}
\end{SCfigure}

Asteroseismology would provide a unique solution to these problems. However, owing to their low luminosities, oscillation amplitudes in dwarfs cooler than the Sun are small and difficult to detect, even with high-precision photometry from \kep. To date, only a handful of stars cooler than the Sun have detected oscillations, and none cooler than 5000\,K (Figure \ref{cooldwarfs}).

A solution is to perform asteroseismology using radial velocities (RV), which are less affected by stellar granulation noise than photometry \citep{harvey88} and thus allow higher S/N detections in cool stars. Since there are no stellar noise sources on timescales shorter than oscillations, dedicated ground-based observations using small telescopes such as the Stellar Observations Network Group \citep[SONG,][]{grundahl17} can build up sufficient S/N to detect oscillations in cool dwarfs. Additionally, high-precision Doppler spectrographs on large telescopes such as the Keck Planet Finder \citep[KPF,][]{gibson16} allow high cadence RVs for a selected number of high priority stars.  Simulations demonstrate that a combination of SONG and KPF would allow the systematic application of asteroseismology in stars cooler than the Sun (Figure \ref{cooldwarfs}). \textbf{Establishing ground-based radial-velocity networks for asteroseismology in the US through SONG nodes and investments in fast, high-precision Doppler spectrographs on large telescopes will enable the first extension of asteroseismology into the regime of cool dwarfs, tackling some of the most severe problems in our understanding of stellar evolution.}

\section{Conclusions}

\noindent
Asteroseismology is the only observational tool in astronomy that can probe the interiors of stars, and a benchmark method for deriving fundamental stellar properties. The field will continue to flourish over the coming decade, and we anticipate the following key science questions and necessary facilities to continue the asteroseismology revolution into the 2020's:

\begin{list}{\labelitemi}{\leftmargin=1em}
\setlength{\itemsep}{2pt}%
\setlength{\parskip}{2pt}
    \item Space-Based Missions such as \tess\ and PLATO will provide asteroseismic data for thousands of solar-type stars. To use these data to address critical issues in stellar interior physics such as convection, rotation and magnetic fields, independent fundamental measurements of stellar properties from optical long-baseline interferometry (\textbf{CHARA, NPOI and MROI}) are required.
    
    \item The reach of galactic archeology using oscillating red giants will dramatically increase over the coming decade, thanks to the wide coverage of space-based missions such as \textbf{\textit{TESS}}, \textbf{PLATO} and \textbf{WFIRST}. Next-generation spectroscopic surveys such as \textbf{SDSS-V} and \textbf{MSE} are essential to provide spectroscopy for the millions of stars with high-precision space-based photometry.
    
    \item Ground-based radial velocities will enable the systematic extension of asteroseismology to dwarfs cooler than the Sun, which are plagued by systematic errors in stellar models. US participation in ground-based networks such as \textbf{SONG} and investments in high-precision RV spectrographs on large telescopes such as \textbf{KPF} will be critical to push asteroseismology into this new frontier.
    
\end{list}

\pagebreak

\bibliographystyle{aasjournal}
\bibliography{references}

\end{document}